\documentstyle[preprint,eqsecnum,aps,axodraw]{revtex}
\def\be{\begin{equation}}
\def\ee{\end{equation}}
\def\bea{\begin{eqnarray}}
\def\eea{\end{eqnarray}}
\def\beas{\begin{eqnarray*}}
\def\eeas{\end{eqnarray*}}

\def\ep{\epsilon}

\def\d{\dagger}
\def\cx{c_x}
\def\cy{c_y}
\begin{document}
\draft
\title{A Study of Light Mesons on the Transverse Lattice}
\author{Matthias Burkardt and Sudip K. Seal}
\address{Department of Physics\\
New Mexico State University\\
Las Cruces, NM 88003-0001\\U.S.A.}
\maketitle
\begin{abstract}
We present results from a study of light meson spectra and structure
obtained within the
framework of light-front QCD formulated on
a transverse lattice. We discuss how imposing Lorentz covariance conditions
on meson dispersion relations allows determination of parameters in the
transverse lattice Hamiltonian. The pion distribution amplitude
obtained in this framework is rather close to its asymptotic shape.
\end{abstract}
%\pacs{Valid PACS appear here}
%{\tt$\backslash$\string pacs\{\}} should always be input,
%even if empty.}
\narrowtext
\section{Introduction}
Many high energy scattering experiments probe hadron structure very close to
the light-cone. Hence correlation functions probed in such experiments have
a particularly simple physical interpretation in the light-front (LF)
framework (or infinite momentum frame), i.e. in a Hamiltonian framework
where $x^+\equiv x^0+x^3$ is `time'.
For example, parton distribution functions measured in
deep-inelastic scattering experiments provide information about the
LF momentum distribution of quarks in the target hadron.
Off-forward parton distributions, probed in deeply virtual Compton
scattering,
have the interpretation of matrix elements of the LF momentum density
operator
between states of unequal momenta.
Furthermore, the pion distribution amplitude, relevant for example for the
asymptotic pion form factor, can be related to the LF wavefunction for the
pion in the $q\bar{q}$ Fock component when $q$ and $\bar{q}$ have the same
$\perp$ position.

Even though these important observables have their most physical
interpretation
on the LF, it is, at least in principle, possible to calculate them in any
approach to QCD. However, what distinguishes the LF framework from all other
formulation of QCD is that the above observables have a very direct and
physical
connection to the microscopic degrees of freedom in terms of which the
Hamiltonian is constructed. Because of this unique feature, it should be
much easier in this framework to gain a physical understanding between
experiment and phenomenology on the one hand and the
underlying QCD dynamics on the other.

Other motivations to study QCD formulated on the LF derive from the fact
that the complexity of the vacuum seemingly shifts from the states
to the operators in this framework \cite{mb:adv}. This results in a
separation between the physics of the vacuum and the parton structure of
hadrons which implies for example that a constituent picture of hadrons has
a chance to make sense \cite{wilson}.

Of course, just like in any other approach to QCD, it is necessary to
regularize
both UV and IR divergences before one can even attempt to perform
nonperturbative calculations.
The transverse lattice \cite{bardeen} is an attempt to combine advantages
of the LF and lattice formulations of QCD.
In this approach to LF-QCD the time and one space direction (say $x^3$)
are kept continuous, while the two `transverse' directions
${\bf x}_\perp \equiv (x^1,x^2)$ are discretized.
Keeping the time and $x^3$ directions continuous has the advantage of
preserving manifest boost invariance for boosts in the $x^3$ direction.
Furthermore, since $x^\pm = x^0 \pm x^3$ also remain continuous,
this formulation still allows a canonical LF Hamiltonian approach.
On the other hand, working on a position space lattice in the transverse
direction allows one to introduce a gauge invariant cutoff on
$\perp$ momenta --- in a manner that is similar to Euclidean or
Hamiltonian lattice gauge theory.

In summary, the LF formulation has the advantage of utilizing degrees of
freedom that are very physical since many high-energy scattering observables
(such as deep-inelastic scattering cross sections) have very simple
and intuitive interpretations as equal LF-time ($x^+$) correlation
functions.
Using a gauge invariant (position space-) lattice cutoff in the $\perp$
direction within the LF framework has the advantage of being able to avoid
the notorious $1/k^+$ divergences from the gauge field in LF-gauge which
plague many other Hamiltonian  LF approaches to QCD \cite{mb:korea}.

The hybrid treatment (continuous versus discrete) of the longitudinal/transverse
directions implies an analogous hybrid treatment of the longitudinal versus transverse
gauge field: the longitudinal gauge field degrees of freedom
are the non-compact $A^\mu$ while the transverse gauge degrees of freedom are
compact link-fields. Each of these degrees of freedom depend on two
continuous
($x^\pm$) and two discrete (${\bf n}_\perp$) space-time variables, i.e. from
a formal point of view the canonical
transverse lattice formulation is equivalent to a
large number of coupled $1+1$ dimensional gauge theories
(the longitudinal gauge fields at each ${\bf n}_\perp$) coupled to nonlinear
$\sigma$ model degrees of freedom (the link fields) \cite{paul}.

For a variety of reasons it is advantageous to work with transverse gauge
degrees
of freedom that are general matrix fields rather than $U \in SU(N_C)$.
First of all, we would like to work at a cutoff scale which is small
(in momentum scale) since only then do we have a chance to find low lying
hadrons that are simple (i.e. contain only few constituents). If one wants
to work on a very coarse lattice, it is useful to introduce
smeared
or averaged degrees of freedom. Upon averaging over neighboring `chains' of
$SU(N_C)$ fields one obtains degrees of freedom which
still transform in the same way as the original $SU(N_C)$ degrees
of
freedom under gauge transformations but are general matrix degrees of
freedom no longer obeying $U^\dagger U=1$ and $det(U)=1$. The price that one has
to pay for introducing these smeared degrees of freedom are more complicated
interactions. The advantage is that low lying hadrons can be described in
a Fock expansion (this has been confirmed by calculations of the static
quark-antiquark potential \cite{mb:bob} and glueball spectra \cite{brett}).

Another important advantage of this `color-dielectric' approach is that it
is much easier to construct a Fock expansion of states out of general linear
matrix fields than out of fields that are subject to non-linear $SU(N_C)$
constraints.

In the color-dielectric approach the complexity
is shifted from the states to the Hamiltonian: In principle, there exists an
exact prescription for the transformation from one set of degrees of freedom
(here $U$'s) to blocked degrees of freedom $M\equiv \sum_{av} \prod_i U_i$
\be
e^{-S_{eff.}(M)} = \int \left[dU\right] e^{-S_{can.}(U)}
\delta \left(M- \sum_{av}\prod_i U_i\right).
\ee
The problem with this prescription is that $S_{eff.}$ is not only very
difficult to determine directly, but in general also contains arbitrarily
complicated interactions.

A much more practical approach towards determining the effective interaction
among the link fields nonperturbatively is the use of Lorentz invariance.
This strategy has been used in a systematic study of glueball masses
in Ref. \cite{brett},
where more details can be found regarding the effective interaction.
One starts by making the most general ansatz for the effective interaction
which is invariant under those symmetries of QCD that are not broken by the
$\perp$ lattice. This still leaves an infinite number of possible terms and
for practical reasons, only terms up to fourth order  in the fields and
only local (in the $\perp$ direction) terms have been included in the
Ref. \cite{brett}. The coefficients of the remaining terms are then fitted
to maximize Lorentz covariance for physical observables, such as the
$Q\bar{Q}$ potential (rotational invariance!) and covariance of the
glueball dispersion relation. It should be emphasized that these are first
principle calculations in the sense that the only phenomenological input
parameter is the overall mass scale (which can for example be taken to be
the lowest glueball mass or the string tension). The only other input that
is used is the requirement of Lorentz invariance.

The numerical results from Refs. \cite{brett,mb:bob} within this approach
are very encouraging:
\begin{itemize}
\item with only a few parameters, approximate Lorentz invariance could be
achieved for a relatively large number of glueball dispersion relations
simultaneously \cite{brett} as well as for the $Q\bar{Q}$ potential
\item the glueball spectrum that was obtained numerically on the $\perp$
lattice for $N_C\rightarrow \infty$ is consistent with Euclidean Monte
Carlo lattice gauge theory calculations performed at finite $N_C$ and
extrapolated to $N_C\rightarrow \infty$ .
\end{itemize}
For further details on these very interesting results,  the reader is
referred
to Refs. \cite{brett,mb:bob} and references therein.
In this paper, we are applying this program to mesons on
the transverse lattice within the femtoworm approximation \cite{mb:hala}.
This Hamiltonian has already been analyzed in Ref. \cite{mb:hala} for mesons
with ${\vec k}_\perp=0$. In this work, as well as in a related independent paper
\cite{sd:hd}, the work from Ref. \cite{mb:hala} is generalized to mesons
with nonzero transverse momenta. This generalization allows one to address
the issue of Lorentz invariance of meson spectra. There are significant
differences in the general approach between our work and Ref.
\cite{sd:hd}. First of all, while we employ continuous basis function
techniques to solve the longitudinal dynamics, Refs. \cite{sd:hd,sd:meson} 
use discrete light-cone quantization (DLCQ). Secondly, there is also 
a difference in the procedure: both this work and Ref. \cite{sd:meson}
use a longitudinal
gauge coupling taken from pure glue calculations, while Ref. 
\cite{sd:hd} refits this parameter. 
The latter difference has important consequences which we discuss
in detail in section III. Our results confirm some of the results
presented in Ref. \cite{sd:meson},  which serves as a non-trivial test for
both approaches. However, the main motivation for our work lies in a
different direction. Our goal was to understand to what extent the
requirement of imposing Lorentz invariance sufficiently constrains the
parameters of the Hamiltonian first studied in Ref. \cite{mb:hala} and also
to investigate whether this Hamiltonian and the femtoworm approximation are
able to yield  Lorentz invariant results.  The paper is organized as
follows.  We briefly describe the Hamiltonian, the approximations used to
solve it and some of it's subtle properties in section II.  In section III, we
provide a perturbative analysis of the $\pi-\rho$ splitting. The results
from this section will be of importance for understanding the
non-perturbative numerical results presented in section IV. This is followed by some
comments and comparisons of this work with another concurrent work in section V and
a brief summary in section VI.

\section{The Hamiltonian}
\begin{figure}
\begin{picture}(0,110)(300,10)
\put(400,62){\circle{5}}
\put(402,61){\line(1,0){45}}
\put(445,62){\circle*{5}}
\put(397,50){n}
\put(436,50){n+1}
\put(412,65){+$\vec{e}_x$}

\put(480,62){\circle{5}}
\put(479.5,65){\line(0,1){42}}
\put(480,107){\circle*{5}}
\put(487,60){n}
\put(487,105){n+1}
\put(487,83){+$\vec{e}_y$}

\put(525,62){\circle*{5}}
\put(523,61){\line(1,0){45}}
\put(570,62){\circle{5}}
\put(518,50){n-1}
\put(568,50){n}
\put(540,65){-$\vec{e}_x$}

\put(600,62){\circle{5}}
\put(599.5,17){\line(0,1){43}}
\put(600,17){\circle*{5}}
\put(607,60){n}
\put(607,15){n-1}
\put(605,37){-$\vec{e}_y$}
\end{picture}
\caption{Four possible orientations of the 3-particle states.}
\label{fig:orient}
\end{figure}
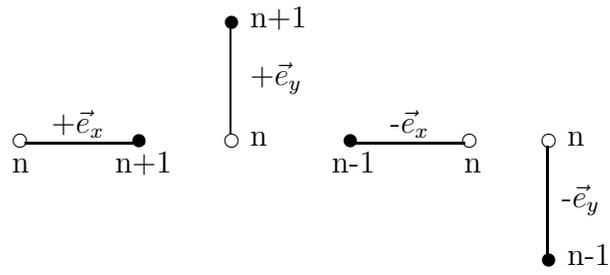

The transverse lattice Hamiltonian for Wilson fermions
that forms the basis of our work and Ref. \cite{sd:meson} was
first constructed in Ref. \cite{mb:hala}, and we refer the reader to 
these works for more details. 
Here we restrict ourselves to a more qualitative
description of the degrees of freedom that enter the Hamiltonian and we
follow up later with more specifics when we discuss the integral equations
describing mesons in the femtoworm approximation.  The degrees of freedom
that enter the Hamiltonian are quark and antiquark creation operators at
each site and link-fields (which, in the spirit of the color-dielectric
approach \cite{brett}, we take as general  matrix fields) on the transverse
links connecting the sites. For simplicity, we work in the large $N_C$ 
limit. The first approximation that we use is a
light front Tamm-Dancoff truncation of the Fock state expansion upto and
including the 3-particle sector. Upon reminding oneself that the fermion
degrees of freedom (quarks and antiquarks)
always occupy the lattice (transverse) sites and the gauge degrees of freedom
(link fields) reside
on the lattice spacings, it is easy to understand the following two direct
consequences of the
aforementioned truncation:
\begin{itemize}
\item The 2-particle states consist of a quark and an antiquark sitting on
the same lattice site.
\item The 3-particle states consist of a quark and an antiquark separated by
{\bf at most} one link field.
\end{itemize}
Since link fields always reside on the lattice spacings as demanded by
gauge invariance, the 3-particle Fock space is therefore spanned by all
different configurations of a
quark and an antiquark separated by one link field. This dumb-bell shaped
configuration of the
3-sector naturally requires one to introduce an index to specify it's
transverse orientation. We do so
as shown in Fig. \ref{fig:orient}.
%\newpage

In keeping with the truncated Fock space, one can therefore write the meson
state vector as:
\be
\mid meson\rangle = c_2\mid q\bar{q}\rangle + c_3\mid q\bar{q}g\rangle
\ee
The total angular momentum of the fermions in the 2-sector is contributed
wholly by their respective
spins since there is no contribution from orbital angular momentum when they
are both on the same site. On the other hand, the total angular momentum of
the
3-particle configuration is the sum of both their spins as well as the
orbital angular momentum typical
of dumb-bell shaped systems. To characterize a 2-particle state, one needs
the spins of the two quarks
along with their momenta, both longitudinal as well as transverse.
Additionally, a 3-particle state
will require the direction of it's configuration (see Fig. \ref{fig:orient})
in the
discretized {\it xy} directions. The Hamiltonian, however, remains invariant
when translated in
multiples of the transverse lattice spacing. Consequently, the transverse
momenta of the mesons are conserved (modulo $\pi$). 
This allows one to solve the light front Hamiltonian equation,
$H_{LF}\mid\psi\rangle=P^-\mid\psi\rangle$, for fixed values of the
transverse momentum, $k_\perp$.
For a given value of $k_\perp$ we therefore make the following ansatz:

\bea
\mid
q\bar{q}\rangle&=&\sum_{\vec{n}_\perp}e^{i\vec{k}_\perp.\vec{n}_\perp}\int_0
^1dx
b^\dagger_{s_1}(x,\vec{n}_\perp) \nonumber \\
& & d^\dagger_{s_2}(1-x,\vec{n}_\perp)\psi_{s_1,s_2}(x)\mid0\rangle
\label{eq:2.2}
\eea

\bea
\mid
q\bar{q}g\rangle&=&\sum_{\vec{n}_\perp}e^{i\vec{k}_\perp.\left(\vec{n}_\perp
+\frac{\vec{e}_i}
{2}\right)}
 \nonumber \\
&
&\int_0^1dx\int_0^{1-x}dyb^\dagger_{s_1}(x,\vec{n}_\perp)d^\dagger_{s_2}(y,\vec{
n}_\perp+\vec{e}_i)
 \nonumber \\
& &a^\dagger(1-x-y)\psi_{s_1,s_2}(x,y,\vec{e}_i,)\mid0\rangle
\label{eq:2.3}
\eea
where the spin degrees of freedom, $s_1,s_2\in(\uparrow,\downarrow)$ and
the orbital degrees of freedom, $\vec{e}_i\in(\pm\vec{e}_x,\pm\vec{e}_y)$.
In the above, $\psi_{s_1,s_2}(x)$ and $\psi_{s_1,s_2}(x,y,\vec{e}_i)$ are
the wavefunction amplitudes in the 2-particle and
3-particle sectors respectively. It is a straight forward though tedious
procedure to show that
the QCD Hamiltonian with the above state vectors in the large $N_c$ limit
and under the light front gauge
yields the following integral equations in terms
of the wavefunction amplitudes in the 2-particle
and 3-particle sectors respectively:

\bea
H_{22}\psi_{s_1,s_2}(x)&=&\left(\frac{m_q^2-1}{x}+\frac{m_{\bar{q}}^2-1}{1-x}\right)
\psi_{s_1,s_2}(x)\nonumber \\
&&-\int^1_0dx'\frac{\psi_{s_1,s_2}(x')}{(x-x')^2}
\label{eq:5}
\eea

\bea
H_{33}&&\psi_{s_1,s_2}(x,y,\vec{e}_i)=\left(\frac{m_q^2-1}{x}+\frac{m_{\bar{q}}^2-1}
{y}+\frac{m_U^2}{1-x-y}\right)
\nonumber \\
&&\psi_{s_1,s_2}(x,y,\vec{e}_i)
-\int^{1-y}_0dx'\frac{\psi_{s_1,s_2}(x',y,\vec{e}_i)}{(x-x')^2}\frac{x_U+x'_
U}{2\sqrt{x_Ux'_U}}\nonumber \\
&& -\int^{1-x}_0dy'\frac{\psi_{s_1,s_2}(x,y',\vec{e}_i)}{(y-y')^2}\frac{y_U+y
'_U}{2\sqrt{y_Uy'_U}}\nonumber \\
\label{eq:6}
\eea
In Eq. (\ref{eq:5}), the first two terms on the right hand side are the kinetic
energies of the quark and the
anti-quark while the third term describes the Coulomb interaction between
them. In Eq. (\ref{eq:6}), the
first three terms on the right hand side are the kinetic energies of the
quark, anti-quark and the link
fields, the fourth term describes the Coulomb
interaction  between the quark(antiquark) and the link field and the fifth
term describes the Coulomb
interaction between the antiquark(quark) and the link field. It should be
noted that a Coulomb
interaction term between the quark and antiquark in the 3-particle sector
does not appear since it is disallowed under the large
$N_c$ limit. Three remarks are in order at this point.
First, $H_{22}$ and $H_{33}$, i.e. the terms in the Hamiltonian that are
diagonal in Fock space are independent of the quark spin and the 
orientation of the link. Second,
the $\psi$'s in Eqs. (\ref{eq:5}) and (\ref{eq:6})
still depend implicitly on $k_\perp$ because although $k_\perp$ is
conserved, $k_\perp$ still enters
the integral equations as a parameter. Finally, Eqs. (\ref{eq:5})  and 
(\ref{eq:6}) have been
rescaled in a manner such that
$G^2=1$.

The natural question that arises at this point is: How do the 2-sector and
the 3-sector mix with each
other. This brings us to the issue of propagation of the mesons on the
lattice. Qualitatively, the
way this
takes place is as follows: A $q$($\bar{q}$) hops to a neighboring site with
the emission of a
link field (2-sector $\rightarrow$ 3-sector). The $\bar{q}$($q$) sitting on
the
original site then jumps to join the $q$($\bar{q}$) by absorbing the link
field
(3-sector $\rightarrow$ 2-sector). Since this propagation resembles that of
an inchworm on a scale of
a femtometer, we were inspired to called this the {\it femtoworm
approximation}. The hopping of a quark or
an antiquark from one site to the next can cause it to either flip its spin
or not. Accordingly, we introduce two distinct couplings between the fermion and
the link degrees of freedom. In this work, we label the spin-flip coupling as
$m_v$ and the non spin-flip coupling as $m_r$. The corresponding interaction terms 
are presented in details later in this section.
The natural question that now arises from the above is: Is it possible for a
link field that is
emitted by a quark(antiquark) through the $m_r$ coupling to be absorbed by
an antiquark(quark)
through the $m_v$ term, in other words, `mixed' hopping ?  The answer to this
is `No'. 
To understand the hopping interactions better one needs to analyze the
rotational properties of the Hamiltonian more closely.
For ${\bf k}_\perp=0$,
the $\perp$ lattice Hamiltonian is invariant under rotations around the
$z-axis$ by multiples of
$\pi/2$, giving rise to conservation of total angular momentum modulo $4$.
Since $J_z=S_z$ in the 2
particle sector and $S_z$ can only assume the values $-1$, $0$, and $1$,
$S_z$  in the 2 particle
sector is conserved. Since mixed hopping would change $S_z$ by one unit,
this means that the sum  of all contributions from mixed hopping must add up
to zero.
The observation that helicity conservation (modulo 4) for
${\vec k}_\perp=0$ prohibits interference between the two hopping terms was
already made in Ref. \cite{mb:hala}. 
%Here we would like to point out that
%even for ${\bf k}_\perp \neq {\bf 0}$ such an interference is not possible.
%Of course, in this case the argument is a little more complicated  since
%rotations also change the direction of ${\bf k}_\perp$.  Nevertheless, both
%the $\perp$ lattice Hamiltonian as well as ${\bf k}_\perp$
%are invariant under   
%a sequence consisting of a rotation by $\pi$ around the $z$-axis followed by
%a $\perp$ reflection on the $x$-axis and then a $\perp$ reflection on the
%$y$ axis $P_yP_xR_{180}$.
%As a result, $J_z$ is conserved modulo $2$ which still rules out mixing
%between the $m_r$-term and the $m_v$-term.

The interaction of the 2-particle sector with the 3-particle sector will
therefore have four different
possibilities:
\begin{itemize}
\item a quark(or an antiquark) in the 2-sector emits a link through the $m_v$
term and flips it's spin.
\item a quark(or an antiquark) in the 2-sector emits a link through the $m_r$ 
term and does not flip it's spin.
\end{itemize}
Of course, the interaction to go from the 3-sector to the 2-sector has the
opposite effect, i.e,
\begin{itemize}
\item a quark(or an antiquark) in the 3-sector absorbs a link through the $m_v$
term and flips it's spin.
\item a quark(or an antiquark) in the 3-sector absorbs a link through the $m_r$ 
term and does not flip it's spin.
\end{itemize}
In order to understand the spin dynamics due to the hopping terms as described above we present below
the interaction Hamiltonian in all its glory.

Spin flip hopping in the $x$-direction:
\bea
\delta H_x^{flip}&=& \\
&-&i\left(\frac{m_v}{p^+_f}-\frac{m_v}{p^+_i}\right)\sum_{\vec{n}_\perp}
\left[ \right.
b^\d_\uparrow(\vec{n}_\perp,p_f^+)b_\downarrow(\vec{n}_\perp+\vec{e}_x,p_i^+) \nonumber \\
& &\quad \quad \quad \quad \quad \quad \quad \quad
-b^\d_\downarrow(\vec{n}_\perp,p_f^+)b_\uparrow(\vec{n}_\perp+\vec{e}_x,p_i^+) \nonumber\\
& &\quad \quad \quad \quad \quad \quad \quad \quad
-b^\d_\uparrow(\vec{n}_\perp,p_f^+)b_\downarrow(\vec{n}_\perp-\vec{e}_x,p_i^+) \nonumber\\
& &\quad \quad \quad \quad \quad \quad \quad \quad
+b^\d_\downarrow(\vec{n}_\perp,p_f^+)b_\uparrow(\vec{n}_\perp-\vec{e}_x,p_i^+) \nonumber\\
& &\quad \quad \quad \quad \quad \quad \quad \quad
-d^\d_\uparrow(\vec{n}_\perp,p_f^+)d_\downarrow(\vec{n}_\perp+\vec{e}_x,p_i^+) \nonumber\\
& &\quad \quad \quad \quad \quad \quad \quad \quad
+d^\d_\downarrow(\vec{n}_\perp,p_f^+)d_\uparrow(\vec{n}_\perp+\vec{e}_x,p_i^+) \nonumber\\
& &\quad \quad \quad \quad \quad \quad \quad \quad
+d^\d_\uparrow(\vec{n}_\perp,p_f^+)d_\downarrow(\vec{n}_\perp-\vec{e}_x,p_i^+) \nonumber\\
& &\quad \quad \quad \quad \quad \quad \quad \quad
-d^\d_\downarrow(\vec{n}_\perp,p_f^+)d_\uparrow(\vec{n}_\perp-\vec{e}_x,p_i^+)  
\left.\right]\nonumber
\eea

Spin flip hopping in the $y$-direction:
\bea
\delta H_y^{flip}&=& \\
&-&\left(\frac{m_v}{p^+_f}-\frac{m_v}{p^+_i}\right)\sum_{\vec{n}_\perp}
\left[ \right.
b^\d_\uparrow(\vec{n}_\perp,p_f^+)b_\downarrow(\vec{n}_\perp+\vec{e}_y,p_i^+) \nonumber \\
& &\quad \quad \quad \quad \quad \quad \quad \quad
+b^\d_\downarrow(\vec{n}_\perp,p_f^+)b_\uparrow(\vec{n}_\perp+\vec{e}_y,p_i^+) \nonumber\\
& &\quad \quad \quad \quad \quad \quad \quad \quad
-b^\d_\uparrow(\vec{n}_\perp,p_f^+)b_\downarrow(\vec{n}_\perp-\vec{e}_y,p_i^+) \nonumber\\
& &\quad \quad \quad \quad \quad \quad \quad \quad
-b^\d_\downarrow(\vec{n}_\perp,p_f^+)b_\uparrow(\vec{n}_\perp-\vec{e}_y,p_i^+) \nonumber\\
& &\quad \quad \quad \quad \quad \quad \quad \quad
-d^\d_\uparrow(\vec{n}_\perp,p_f^+)d_\downarrow(\vec{n}_\perp+\vec{e}_y,p_i^+) \nonumber\\
& &\quad \quad \quad \quad \quad \quad \quad \quad
-d^\d_\downarrow(\vec{n}_\perp,p_f^+)d_\uparrow(\vec{n}_\perp+\vec{e}_y,p_i^+) \nonumber\\
& &\quad \quad \quad \quad \quad \quad \quad \quad
+d^\d_\uparrow(\vec{n}_\perp,p_f^+)d_\downarrow(\vec{n}_\perp-\vec{e}_y,p_i^+) \nonumber\\
& &\quad \quad \quad \quad \quad \quad \quad \quad
+d^\d_\downarrow(\vec{n}_\perp,p_f^+)d_\uparrow(\vec{n}_\perp-\vec{e}_y,p_i^+)  
\left.\right]\nonumber
\eea

Hopping without spin flip:
\bea
\delta H^{no-flip}&=& \\
&-&\left(\frac{m_r}{p^+_f}+\frac{m_r}{p^+_i}\right)\sum_{\vec{n}_\perp}
\left[ \right.
b^\d_\uparrow(\vec{n}_\perp,p_f^+)b_\uparrow(\vec{n}_\perp+\vec{e}_x,p_i^+) \nonumber \\
& & \quad \quad \quad \quad \quad \quad \quad
+b^\d_\uparrow(\vec{n}_\perp,p_f^+)b_\uparrow(\vec{n}_\perp-\vec{e}_x,p_i^+) \nonumber \\
& & \quad \quad \quad \quad \quad \quad \quad
+b^\d_\downarrow(\vec{n}_\perp,p_f^+)b_\downarrow(\vec{n}_\perp+\vec{e}_x,p_i^+) \nonumber\\
& & \quad \quad \quad \quad \quad \quad \quad
+b^\d_\downarrow(\vec{n}_\perp,p_f^+)b_\downarrow(\vec{n}_\perp-\vec{e}_x,p_i^+) \nonumber\\
& & \quad \quad \quad \quad \quad \quad \quad
+d^\d_\uparrow(\vec{n}_\perp,p_f^+)d_\uparrow(\vec{n}_\perp+\vec{e}_x,p_i^+) \nonumber\\
& & \quad \quad \quad \quad \quad \quad \quad
+d^\d_\uparrow(\vec{n}_\perp,p_f^+)d_\uparrow(\vec{n}_\perp-\vec{e}_x,p_i^+) \nonumber\\
& & \quad \quad \quad \quad \quad \quad \quad
+d^\d_\downarrow(\vec{n}_\perp,p_f^+)d_\downarrow(\vec{n}_\perp+\vec{e}_x,p_i^+) \nonumber\\
& & \quad \quad \quad \quad \quad \quad \quad
+d^\d_\downarrow(\vec{n}_\perp,p_f^+)d_\downarrow(\vec{n}_\perp-\vec{e}_x,p_i^+) \nonumber\\
& & \quad \quad \quad \quad \quad \quad \quad
+b^\d_\uparrow(\vec{n}_\perp,p_f^+)b_\uparrow(\vec{n}_\perp+\vec{e}_y,p_i^+) \nonumber\\
& & \quad \quad \quad \quad \quad \quad \quad
+b^\d_\uparrow(\vec{n}_\perp,p_f^+)b_\uparrow(\vec{n}_\perp-\vec{e}_y,p_i^+) \nonumber\\
& & \quad \quad \quad \quad \quad \quad \quad
+b^\d_\downarrow(\vec{n}_\perp,p_f^+)b_\downarrow(\vec{n}_\perp+\vec{e}_y,p_i^+) \nonumber\\
& & \quad \quad \quad \quad \quad \quad \quad
+b^\d_\downarrow(\vec{n}_\perp,p_f^+)b_\downarrow(\vec{n}_\perp-\vec{e}_y,p_i^+) \nonumber\\
& & \quad \quad \quad \quad \quad \quad \quad
+d^\d_\uparrow(\vec{n}_\perp,p_f^+)d_\uparrow(\vec{n}_\perp+\vec{e}_y,p_i^+) \nonumber\\
& & \quad \quad \quad \quad \quad \quad \quad
+d^\d_\uparrow(\vec{n}_\perp,p_f^+)d_\uparrow(\vec{n}_\perp-\vec{e}_y,p_i^+) \nonumber\\
& & \quad \quad \quad \quad \quad \quad \quad
+d^\d_\downarrow(\vec{n}_\perp,p_f^+)d_\downarrow(\vec{n}_\perp+\vec{e}_y,p_i^+) \nonumber\\
& & \quad \quad \quad \quad \quad \quad \quad
+d^\d_\downarrow(\vec{n}_\perp,p_f^+)d_\downarrow(\vec{n}_\perp-\vec{e}_y,p_i^+)  
\left.\right]\nonumber
\eea
Implicit in the above expressions are link field creation and annihilation 
operators on the link connecting the initial and final state.

The transition matrix elements are obtained by sandwiching the above interactions between the
2-particle and the 3-particle states as written in
Eqs. (\ref{eq:2.2}) and (\ref{eq:2.3}) and
are provided in the table in appendix B. 

The schematic structure of the Hamiltonian therefore looks like the
following:
\bea
H\left[\begin{array}{c}
\psi(x)\\
\psi(x,y)\end{array}\right]=
\left[\begin{array}{cc}
H_{22} & H_{23} \\
H_{32} & H_{33}\end{array}\right]
\left[\begin{array}{c}
\psi(x)\\
\psi(x,y)\end{array}\right] \nonumber\\. \label{eq:Ham}
\eea
where $H_{23}$ and $H_{32}$ involve all the transition matrix elements.

In order to evaluate all the matrix elements,
we expanded the wavefunction amplitudes using a basis of continuous functions as follows:
\be
\psi(x)=\sum_{k=0}^\Lambda kx^{\beta+k}(1-x)^\beta
\ee
\be
\psi(x,y)=\sum_{k=0}^\Lambda\sum_{l=0}^{\Lambda-k}a_{kl}x^{\beta+k}y^{\beta+
l}(1-x-y)^{\beta+\frac{1}{2}}
\ee
where the $\Lambda$ in the Eq. 2.10
and that in Eq. 2.11 need
not be equal.

In the table (see appendix B), $F^r_q$($F^r_{\bar{q}}$) are the
overlap integrals when a quark(antiquark) in the 2-sector goes to the
3-sector by emitting a link field
through the $m_r$ hopping term (the spin non-flip term) and
$f_q$($f_{\bar{q}}$) are the
overlap integrals when a
quark(antiquark) in the 2-sector goes to the 3-sector by emitting a link
field through the $m_v$
hopping term (the spin-flip term). See appendix A for more details
of these overlap integrals.

A factor $\left(\frac{x'}{x}\right)^\ep$ was introduced in the integrand
of the transition matrix elements 
to regularize the infrared (small $x, x'$) divergence
that arises from the self energy of quarks in the case when the quark that
emits a link
field also absorbs it. Of course, all physical observables will have to be
independent of the
value of $\ep$ as will be demonstrated in section IV.

The result of all the above is the following matrix equation for each $k$ and $k,l$:
\bea
H\left[\begin{array}{cc}
N_{k,k'} & 0\\
0 & N_{kl,k'l'}\end{array}\right]
&&\left[\begin{array}{c}
a_{k'}\\
a_{k'l'}\end{array}\right]
= \nonumber \\
&&\left[\begin{array}{cc}
M_{kk'} & M_{k,k'l'} \\
M_{k'l',k} & M_{kl,k'l'}\end{array}\right]
\left[\begin{array}{c}
a_{k'}\\
a_{k'l'}\end{array}\right] \nonumber\\.\label{eq:Ham2}
\eea
where $N_{k,k'}$ and $N_{kl,k'l'}$ are the norm matrices, $M_{kk'}$ and 
$M_{kl,k'l'}$ are the matrix elements in the pure 2-particle and 3-particle sectors, 
respectively and finally $M_{k,k'l'}$ and $M_{k'l',k}$ are the transition matrix elements.
The above matrix equation is then converted into an eigenvalue equation that
yields eigenvectors in
terms of the $a_k$'s and $a_{kl}$'s. 

\begin{figure}
%\begin{Large}
\unitlength1.cm
\begin{picture}(15,7.0)(-11.5,0.5)
\includegraphics{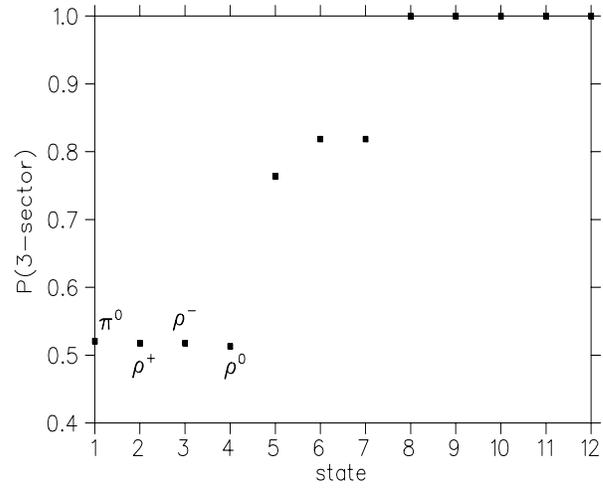}
\end{picture}
%\end{Large}
\caption{Probability of states to be in the 3-particle sector.}
\label{fig:prob}
\end{figure}

Another significant result of this approximation that is worth mentioning at
this point is that
unlike the low lying mesons, the higher excited states are not expected to
show the
correct dispersion. The reason for this is that since the higher states are
very
excited, the quark and the antiquark will be physically farther apart. This
would mean
that these mesons will mostly or always remain in the $q\bar{q}U$ (where $U$
is
the link field) state (see Fig. \ref{fig:prob}) and for it to propagate, it
will need to go to a $q\bar{q}UU$ state or
states with larger number of link fields. But since our approximation does
not allow
states with more than one link field, these excited states simply cannot
propagate.

\section{Perturbative analysis of the $\pi$-$\rho$ splitting}
In order to gain a qualitative understanding about the interplay between
different parameters in the $\perp$ lattice Hamiltonian, it is instructive
to study a simple model, where one treats the admixture of the 3-particle
Fock component to the $\pi$ and $\rho$ as a perturbation.

To $0^{th}$ order, i.e. when the coupling between 2 and 3 particle Fock
component is turned off, there is no spin dependence of the interactions
and the $\pi$ and $\rho$ are degenerate. Likewise, there is no `hopping'
(i.e. $\perp$ propagation) of mesons and thus energies are independent of
${\bf k}_\perp$ giving rise to an infinite $\perp$ lattice spacing
(in physical units).

In the next order we treat the coupling between 2 and 3 particle Fock
components as a perturbation (note that interactions which are diagonal in
the particle
number, such as the confining interaction in the longitudinal direction are
still treated non-perturbatively). As described in section II, there are two
interactions that mix Fock sectors: hopping due to the $m_r$-term (without
helicity flip) and hopping due to the vector coupling $m_v$ (with helicity
flip).
Starting from a basis of `t Hooft eigenstates which are plane waves in the
$\perp$ direction and where the $q\bar{q}$ in the 2 particle Fock component
carry spins $ |\uparrow \uparrow\rangle$,
$ |\uparrow \downarrow\rangle$, $ |\downarrow \uparrow\rangle$,
and $ |\downarrow \downarrow\rangle$ respectively, one thus finds for the
energy in second order perturbation theory to be:

\bea
H = M_0^2 &-& M_{1,r}^2\left(\begin{array}{cccc}
\cx + \cy & 0 & 0 & 0\\
0 & \cx + \cy & 0 & 0\\
0 & 0 & \cx + \cy & 0\\
0 & 0 & 0 & \cx + \cy\end{array}\right) \nonumber\\
&+& M_{1,v}^2\left(\begin{array}{cccc}
0 & 0 & 0 & \cy - \cx\\
0 & 0 & \cx + \cy & 0\\
0 &  \cx + \cy & 0 & 0\\
\cy-\cx  & 0 & 0 & 0\end{array}\right) . \label{eq:M2} \nonumber\\
\eea

Here $M_{1,r}^2$ and $M_{1,v}^2$ are some second order perturbation theory
expressions involving matrix elements between 2 and 3 particle states that
are
eigenstates of the diagonal parts of the Hamiltonian (kinetic + Coulomb),
and $c_i \equiv \cos k_i$.

Several general and important features can be read off from this result.
First of all, and most importantly, Eq. (\ref{eq:M2})
shows that the r-term gives rise to a dispersion relation with the same
$\perp$ speed of light for the $\pi$ and the $\rho$'s,
while the vector interaction breaks that symmetry.
This observation already indicates that it may be desirable to keep the
$r$-term
much larger than the spin-flip term. We will elaborate on this point below.

At $k_x=k_y=0$, the eigenstates of the above Hamiltonian are the
$\rho_{\pm 1}$ i.e.
$|\uparrow \uparrow\rangle$ and
$|\downarrow \downarrow\rangle$,
with $M^2= M^2_{\pm 1}\equiv M_0^2 - M_{1,r}^2$,
the $|\rho_0\rangle\equiv
|\uparrow \downarrow+ \downarrow \uparrow\rangle$,
with  $M^2= M^2_{\pm 1} + M_{1,v}^2$ and
the   $|\pi\rangle\equiv
|\uparrow \downarrow- \downarrow \uparrow\rangle$,
with  $M^2= M^2_{\pm 1} - M_{1,v}^2$
For nonzero $\perp$ momenta, there will in general be mixing
among the $\rho_{+1}$ and the $\rho_{-1}$, but not among the other states
since helicity in the 2-particle Fock sector is still conserved modulo 2.
Expanding around ${\bf k}_\perp=0$, and denoting
$\bar{M}^2\equiv M_0^2-M_{1,r}^2$
one finds to
${\cal O}({\bf k}_\perp^2)$ the following eigenstates and eigenvalues.

\be
\begin{array}{c|c|c}
\mbox{state} & M^2(0) & M^2({\bf k}_\perp^2)-M^2(0)\\ \hline
 & & \\
\uparrow \downarrow -\downarrow \uparrow & \bar{M}^2 -M_{1,v}^2
&
M_{1,r}^2\frac{k_x^2+k_y^2}{2}+M_{1,v}^2\frac{k_x^2+k_y^2}{2}\\
& & \\
\uparrow \downarrow +\downarrow \uparrow & \bar{M}^2+M_{1,v}^2
 &
M_{1,r}^2\frac{k_x^2+k_y^2}{2}-M_{1,v}^2\frac{k_x^2+k_y^2}{2}\\
& & \\
\uparrow \uparrow -\downarrow \downarrow & \bar{M}^2
 &
{M_{1,r}^2}\frac{k_x^2+k_y^2}{2}
-{M_{1,v}^2}\frac{k_x^2-k_y^2}{2}\\
& & \\
\uparrow \uparrow +\downarrow \downarrow & \bar{M}^2
 &{M_{1,r}^2}\frac{k_x^2+k_y^2}{2}
+{M_{1,v}^2}\frac{k_x^2-k_y^2}{2}
\end{array} \label{eq:disp}.
\ee

Eq. (\ref{eq:disp}) illustrates a fundamental dilemma that hampers any
attempt to fully restore Lorentz invariance within the femtoworm
approximation: $M_{1,v}^2$ not only governs the splitting between
the $\pi$ and the (h=0) $\rho$ but is also responsible for violations
of Lorentz invariance among the different helicity states: If one
determines the $\perp$ lattice spacing in physical units for each
meson separately, by demanding that the
$\perp$ speed of light equals $1$, one finds for example
\bea
\left. \frac{1}{a^2_\perp} \right|_\pi &=& M_{1,r}^2 + M_{1,v}^2
\nonumber\\
\left. \frac{1}{a^2_\perp} \right|_{\rho_0} &=& M_{1,r}^2 - M_{1,v}^2
\eea
i.e. increasing the $\pi-\rho$ splitting is typically accompanied by
an increase in Lorentz invariance violation
\be
\left. \frac{1}{a^2_\perp} \right|_\pi -
\left. \frac{1}{a^2_\perp} \right|_\rho = M_{\rho_0}^2 - M_\pi^2 .
\ee
For the $\rho_{\pm 1}$ the breaking is of a similar scale, plus one
also observes an anisotropy in the dispersion relation on the same
scale.

\begin{figure}
%\begin{Large}
\unitlength1.cm
\begin{picture}(15,7.0)(-11.5,0.5)
\includegraphics{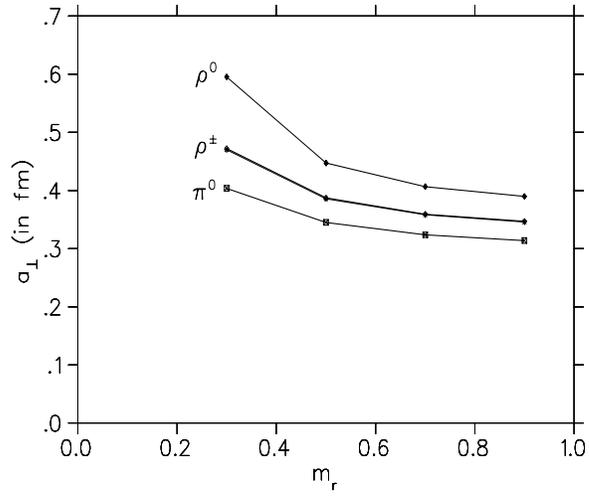}
\end{picture}
%\end{Large}
\caption{Dependence of $a_\perp$ on the non-spin flip coupling.}
\label{fig:m_r}
\end{figure}

Therefore, in order to avoid a large breaking of Lorentz invariance,
it will be necessary that
\be
M_{r,1}^2 \gg M_{\rho_0}^2 - M_\pi^2.
\ee
If one keeps the $\pi-\rho$ splitting fixed at its physical value then
there are two ways to achieve this condition. One possibility is to simply
increase the Yukawa coupling that appears in the $r$-term.
This increase of the $r$-term tends to decrease the $\perp$ lattice spacing
for both $\pi$ and $\rho$'s (see Fig. \ref{fig:m_r}) and in order to achieve
satisfactory
Lorentz invariance (in the sense of uniform $\perp$ lattice spacings)
one needs to make the lattice spacing smaller than the Compton wavelength of
the $\rho$ meson.
However, one cannot make the $r$-term coupling arbitrarily
large because at some point there occurs an instability (tachyonic $M^2$!).
Such instabilities for large coupling are common in the LF formulation
of models with Yukawa coupling and might be
related to a phase transition (similar to the phase transition in $\phi^4$
theory that occurs as the coupling is increased).

Fortunately, there exists another possibility to make these matrix
elements large, without increasing the Yukawa couplings. This derives
from the fact that the hopping interactions are proportional to
$\left(\frac{1}{x} \pm \frac{1}{x^\prime}\right)\frac{1}{x-x^\prime}$, where
$x$ ($x^\prime$) are the momenta of the active quark before(after) the
hopping. Because of the singularity as $x$, $x^\prime \rightarrow 0$, matrix
elements of the hopping terms are greatly enhanced if
the unperturbed wave functions are large near $x=0$ and $x=1$.
Since the unperturbed wave functions in the 2 particle Fock component vanish
like $x^\beta$ near $x=0$, where $\beta \propto \frac{G}{m_q}$,
matrix elements of the hopping interaction become very large when
one makes $\frac{G^2}{m_q^2}$ very large.

Therefore, the larger one chooses $\frac{G^2}{m^2}$, the more one restores
Lorentz invariance of the $\pi$ and $\rho$ dispersion relations because one
can keep the $\pi$-$\rho$ splitting fixed while decreasing the coupling of
the
spin flip interaction. At the same time, keeping the $r$-term interaction
fixed one increases the dominance of the $r$-term contribution in
$\frac{1}{a^2}$ and thus not only reduces the lattice spacing in physical
units, but also obtains dispersion relations for the $\pi$ and the $\rho$'s
that look
more and more similar --- as demanded by Lorentz invariance.
Unfortunately, we are not completely free to pick whatever value of
$\frac{G^2}{m^2_q}$ we like because $m_q^2$ and $G^2$ are largely fixed by
the
center of mass in the $\pi$-$\rho$ system as well as by fitting the physical
string tension in the pure glue sector.

The above analysis may also shed some light on the shape of the
pion distribution amplitude in Ref. \cite{sd:hd}, which is much more flat
(actually, it even exhibits a slight double-hump shape)
compared to our results (see next section). The reason is that the numerical
value for $G^2$ in Ref. \cite{sd:hd} was {\it not} taken from the pure glue
calculation, but was rather used as a free parameter, which was also varied
in order to maximize Lorentz covariance. Although such a {\it first
principle}
approach would normally be desirable, minimizing Lorentz covariance violations
for the $\pi$ and $\rho$ dispersion relations drives $G^2\rightarrow
\infty$.
When the longitudinal gauge coupling $G^2$ is very large compared to other
scales in the problem, the wavefunction tends to become very flat
--- a phenomenon familiar from the 't Hooft equation.
Of course, since a DLCQ cutoff was employed in Ref. \cite{sd:hd}, $G^2$
never really become infinite, but through the extrapolation procedure the
near flat shape for the pion distribution
amplitude
emerged. In a more recent paper \cite{sd:meson}, where Dalley adopted a 
procedure similar to ours and $G^2$ was kept fixed, a pion distribution 
amplitude is obtained that is much more consistent with ours.

\section{Nonperturbative Numerical results}
One cannot expect this Hamiltonian to describe the higher excited states
because of the truncation of the Fock space upto the 3-particle sector. 
This is demonstrated in Fig. \ref{fig:prob} where we show typical results
for the probability to find each state to be in the 3-particle sector.
One finds that whereas the low-lying states have a fairly large probability
to be in the 2-particle sector, it is quite the opposite for 
the higher states. We therefore expect that excited states are much more
strongly affected by the ommission of 4 and more particle states and
consider in the following only the $\pi$ and $\rho$ states in our study 
of Lorentz covariance.

The parameters in the Hamiltonian are:
\begin{itemize}
\item longitudinal gauge coupling
\item $r$-term coupling (hopping without spin flip)
\item spin-flip coupling (hopping with spin flip)
\item kinetic mass (2 particle sector)
\item kinetic mass (3 particle sector). The observables that we studied
showed little dependence on this parameter, so we kept it fixed at a
constituent mass.
\item link field mass. Similar to kinetic mass in 3 particle sector.
Furthermore, demanding Lorentz invariance in the pure glue sector, one finds
a renormalized trajectory. This makes sense since the $\perp$ lattice
scale is unphysical. We keep the link field mass fixed at a value which
yields
relatively small $\perp$ lattice spacing.
\end{itemize}

In the spirit of the color-dielectric approach \cite{brett}, we would like
to use only chiral symmetry (small pion mass), Lorentz invariance and one
length scale (e.g. physical $\rho$ mass or string tension) as the only
input into our calculations.
However, as explained in the previous section, if one uses as input
only the $\pi$ and $\rho$ masses, demanding exact Lorentz invariance
within the framework of the femtoworm approximation
would drive $G^2/m_\rho^2$ to infinity and would thus
give rise to string tensions that are inconsistent with phenomenology.
Likewise, as input only the $\pi$ mass together with Lorentz invariance
would give rise to a degenerate $\pi$-$\rho$ system.
It is thus clear that, as long as one does not go beyond a truncation of
the Fock space above 3 particles ($q$, $\bar{q}$ and one link quantum),
a first principle calculation (where only one mass scale is used as input)
will not be possible. In the following, we will therefore use two
phenomenological scales as input parameters:
the physical value of the $\rho$ mass, as well as the physical string
tension.

Although the dependence of physical parameters on the input parameters is
in general rather complex and non-perturbative, one can understand at a
qualitative level how the input parameters influence the relevant physical
scales: The physical string tension in the longitudinal direction determines
$G^2$ in physical units.

Modulo mass renormalization due to the Yukawa couplings,
the quark mass in the 2-particle Fock component and the gauge coupling are
strongly constrained by fitting the physical string tension (which
determines
$G^2$ and the center of mass of the $\pi$-$\rho$ system.
This leaves us with only the $r$-term and the helicity flip hopping term
couplings as parameters to vary. The helicity
flip term is not only responsible for $\pi$-$\rho$ splitting within our
approximations but also for violations of Lorentz invariance (different
$\perp$ lattice spacings in physical units for different mesons).

\begin{figure}
%\begin{Large}
\unitlength1.cm
\begin{picture}(15.0,6.5)(-11.5,0.9)
\includegraphics{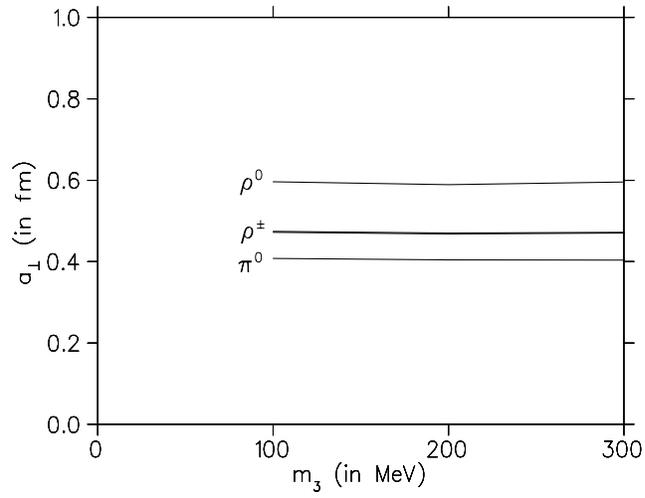}
\end{picture}
%\end{Large}
\vspace{.2in}
\caption{Dependence of the $\perp$ lattice spacings on the masses in the
3-particle Fock component (other parameters are adjusted correspondingly
to keep the physical $\pi$ and $\rho$ masses fixed).}
\label{fig:3sector}
\end{figure}

The observables that we studied, including Lorentz invariance, show
rather little sensitivity \footnote{As long as we kept the other parameters
floating!} to the precise values of the 3-particle sector quark masses
(Fig. \ref{fig:3sector}), which we
thus keep at a value corresponding to a constituent mass (about half the
$\rho$ mass). This leaves us with 4 free parameters.

\begin{figure}
%\begin{Large}
\unitlength1.cm
\begin{picture}(15,7.0)(-11.5,0.5)
\includegraphics{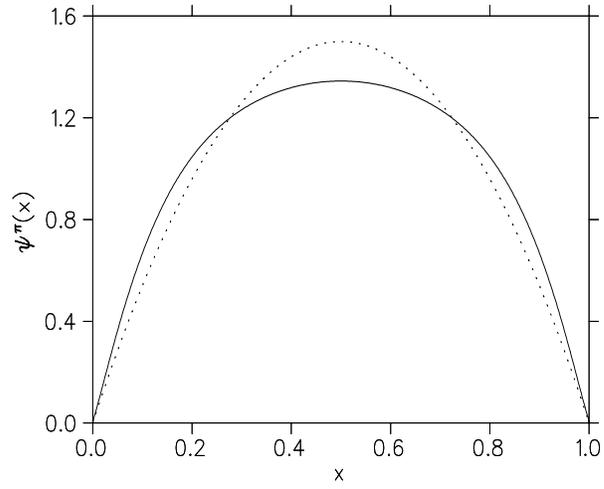}
\end{picture}
%\end{Large}
\caption{Light-cone momentum distribution amplitude for the $\pi$ obtained
on the $\perp$ lattice. The
dashed line is the asymptotic shape.}
\label{fig:pion}
\end{figure}

Using a value of $G^2 \equiv \frac{g^2N_C}{2\pi} \approx
0.4$ $GeV^2$  in physical
units, as determined from the string tension in Ref. \cite{brett} (which is
larger than the one previously used
\cite{mb:hala,sudip}), we were able to produce the physical $\pi$-$\rho$
splitting
with only a relatively small spin-flip coupling. This allows us to chose
the $r$-term large enough so that $r$-term hopping dominates over spin-flip
hopping and therefore violations of Lorentz invariance (as measured by
comparing
quadratic terms in the dispersion relation of mesons in the $\pi$-$\rho$
sector) are only on the order of $20 \%$. The average $\perp$ lattice
spacing
(from the dependence of the energy on ${\bf P}_\perp$) is found to be
$a_\perp \approx 0.5
fm$.
For the $\pi$ wave function, we find a shape (see Fig. \ref{fig:pion}) that
is very close to the asymptotic shape $\phi_{asy}(x)=6x(1-x)$.

\begin{figure}
%\begin{Large}
\unitlength1.cm
\begin{picture}(15,7.0)(-11.5,0.5)
\includegraphics{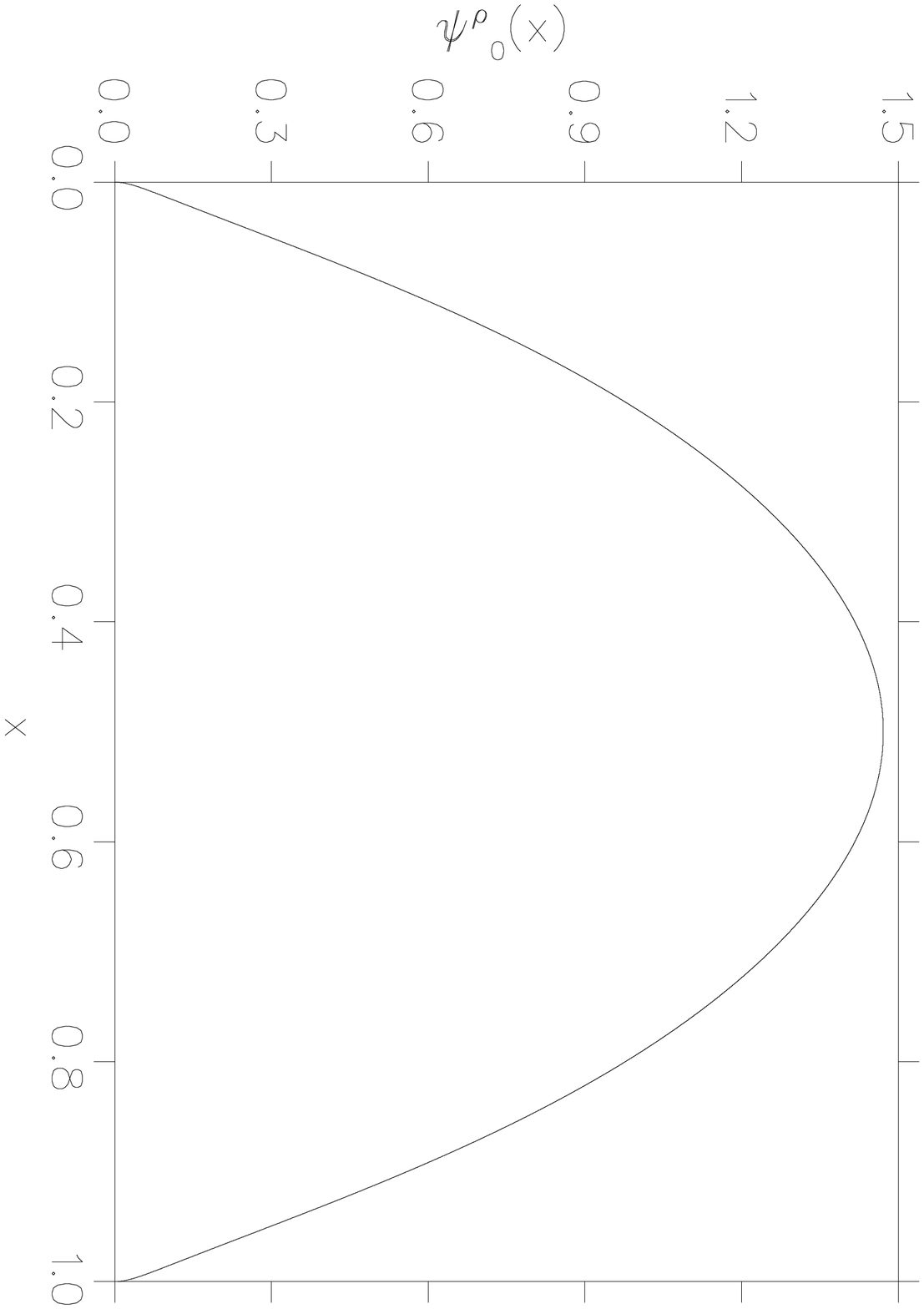}
\end{picture}
%\end{Large}
\caption{Light cone distribution amplitude for $\rho^0$
obtained on the $\perp$ lattice.}
\label{fig:rho0}
\end{figure}

\begin{figure}
%\begin{Large}
\unitlength1.cm
\begin{picture}(15,7.0)(-11.5,0.5)
\includegraphics{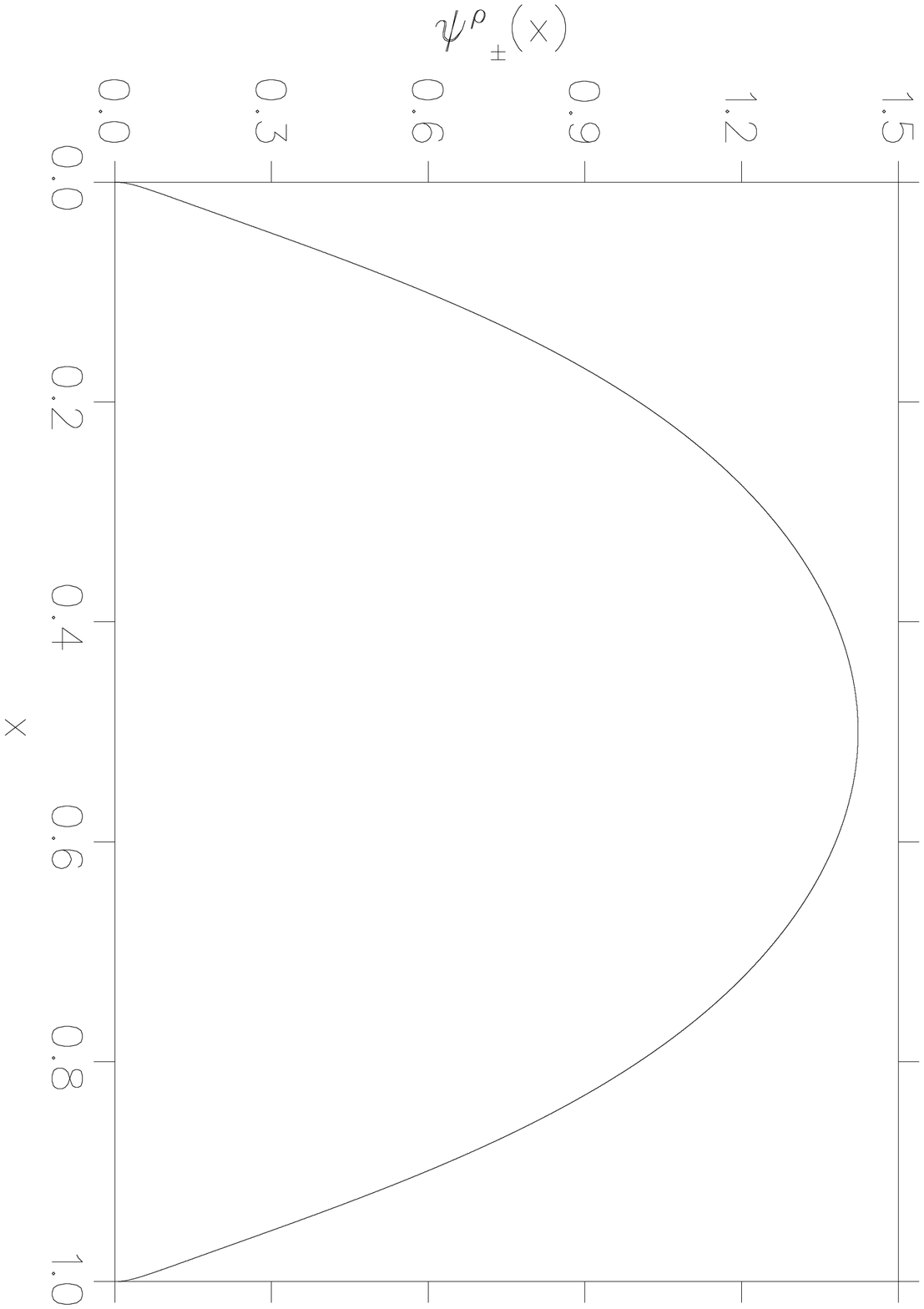}
\end{picture}
%\end{Large}
\caption{Light cone distribution amplitude for $\rho^\pm$
obtained on the $\perp$ lattice.}
\label{fig:rho1}
\end{figure}

\begin{figure}
\unitlength1.cm
\begin{picture}(15,6.0)(-11.5,0.5)
\includegraphics{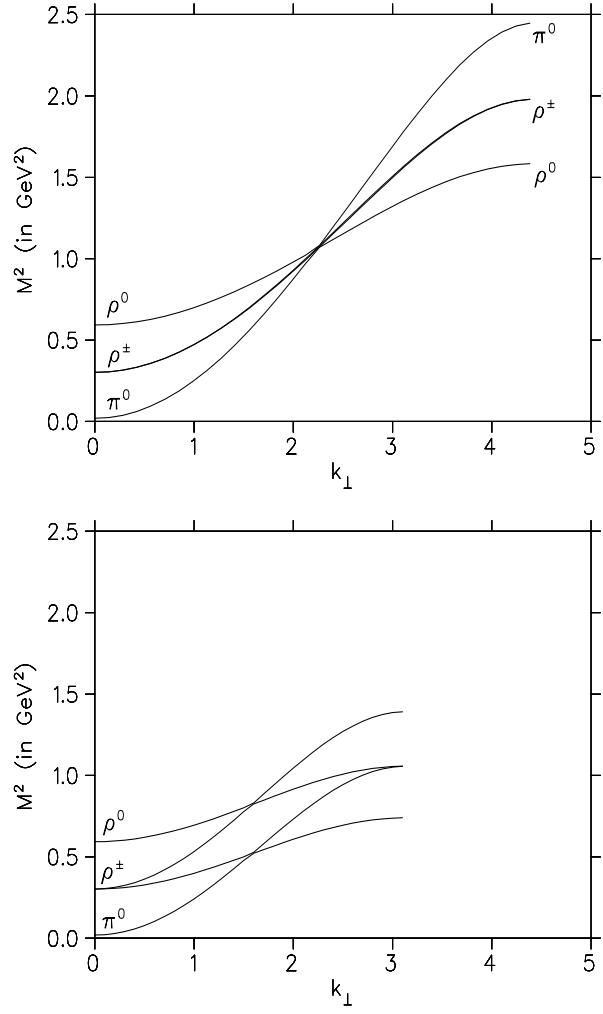}
\end{picture}
\caption{The diagonal(top) and parallel(bottom) dispersion relations for the $\pi$ and $\rho$ mesons.}
\label{fig:bzonep}
\end{figure}

This is surprising if one considers that the lattice spacing is still
relatively large and hence
the momentum scale is still very low. The $\rho$ meson distribution
functions are shown in
Figs. \ref{fig:rho0} and \ref{fig:rho1}.
Notice that the momentum dependence of the $\pi$ dispersion
relation is stronger than the one for the $\rho$ and for the parameters used
in our  calculations this even leads to a level crossing \footnote{Note that
there is no mixing at the crossover point since the $\pi$ and $\rho$ have
opposite C-parity.} The reason for this behavior is very simple. As pointed
out in  Ref. \cite{mb:hala}, for ${\vec k}_\perp=0$ spin-flip hopping  is
attractive  for the $\pi$ and repulsive for the $\rho_0$, while the opposite
is true at the end of the Brillouin zone. More precisely, for $k_x=0$, the
hopping term in the $x$ direction is attractive for the $\pi$ and repulsive
for the $\rho^0$ with the opposite being true for $k_x=\pi$. An analogous
statement holds for the $y$-direction. If there were no $r$-term present,
the dispersion would arise solely from the spin flip term and the $\rho_0$
in the corner of the Brillouin  zone would be degenerate with the $\pi$ at
${\vec k}_\perp=0$ --- a phenomenon which is nothing but a manifestation of
species doubling at the hadronic level. Of course, since we have included an
$r$-term, the $\rho^0$ in the  corner of the Brillouin zone is no longer
degenerate with the $\pi$ at  ${\vec k}_\perp=0$, but the level crossing
remains (see Fig. \ref{fig:bzonep}). 

Similarly one can understand the degeneracy of the $\pi$ and
$\rho^0$ at the edge of the Brillouin zone ${\vec k}_\perp =(\pi,0)$. At that
point, the x-hopping term is repulsive for the $\pi$ and attractive for
the $\rho^0$. The opposite is true for the $y$-hopping  term and the two
contributions cancel one another leaving the $\pi$ and the $\rho^0$
degenerate.
%The diagonal ($k_x=k_y$)  and parallel ($k_x\neq0, k_y=0$) dispersion
%relations
%are shown in Fig. \ref{fig:bzonep}.

From the normalization of the $\pi$ distribution function we were able to
extract a numerical estimate for the $\pi$ decay constant
$f_\pi p^\mu = \langle 0| A^\mu|p\rangle$ in terms of the $+-$ helicity component
of the $\pi$ wavefunction in the two particle Fock component
$\Phi_{+-}(x)$
\be
f_\pi = \frac{2}{a_\perp} \sqrt{\frac{N_C}{\pi}} \int_0^1 dx \psi_{+-}(x)
\approx 300 \, MeV ,
\label{eq:fpi}
\ee
which is much larger than the experimental value $f_\pi^{exp}\approx 93\, MeV$.
We expect that including higher Fock components will 
reduce the size of the two particle Fock component and hence also $f_\pi$.

\begin{figure}
\unitlength1.cm
\begin{picture}(15,6.6)(-11.5,0.5)
\includegraphics{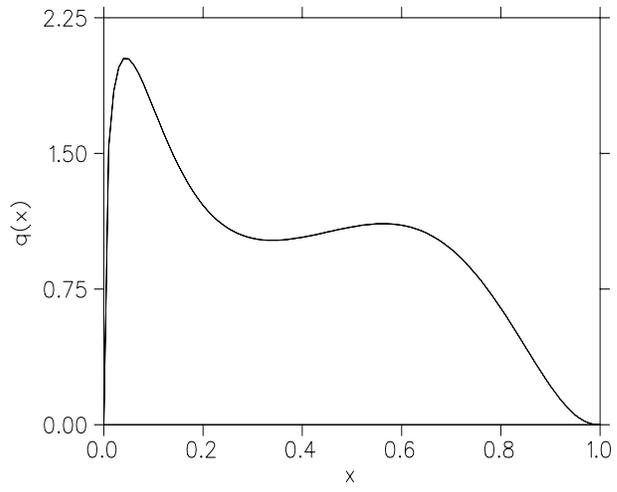}
\end{picture}
\caption{Pion parton distribution function.}
\label{fig:pdf}
\end{figure}

The parton distribution function for the $\pi$ is shown in Fig. \ref{fig:pdf}.
The fact that $q(x)$ goes to zero at small $x$ is an artifact of
our $\ep$-cutoff and we expect a nonzero value for $q(0)$ in
the femtoworm approximation as $\ep\rightarrow 0$.

Finally, we also calculated the electro-magnetic form factor for the $\pi$ for
purely transverse momentum transfers (where overlap formulas can be used)
and we find an rms radius for the $\pi$ of 
$r_\pi^2 \equiv -6\frac{d}{dQ^2} \left.F(Q^2)\right|_{Q=0}
\approx (0.243\, fm)^2$. Obviously, since our lattice spacing is only
about $0.5\, fm$ and since we do not let the quark-antiquark separation
to exceed one lattice unit, one should not expect very good results for
this observable. Since the rms-radius is already poor, displaying the 
form factor is not very meaningful.

\begin{figure}
\unitlength1.cm
\begin{picture}(15,6.6)(-11.5,0.5)
\includegraphics{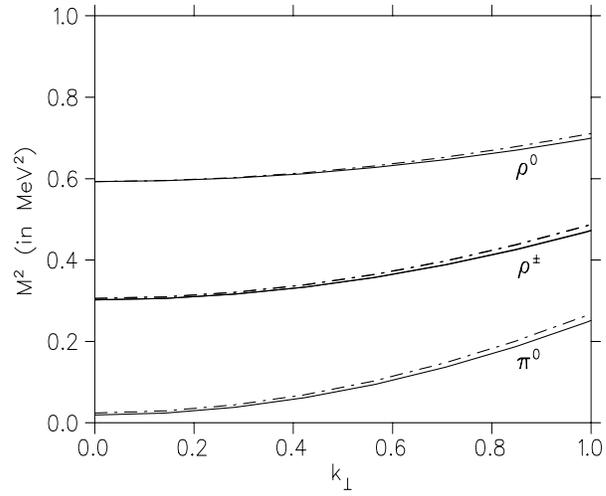}
\end{picture}
\caption{$\ep$ independence of dispersion relations.
The dashed line is for $\ep$ = 0.05 and the continuous line is for 
$\ep$=0.1.}
\label{fig:epsilon}
\end{figure}

As mentioned in section II, physical observables
should be independent of the infrared
cut-off, $\ep$ (as long as the cutoff is small). We verified this
cut-off independence for a variety of observables. As an example, we 
demonstrate in Fig. \ref{fig:epsilon} the cutoff independence of the
dispersion relations for light mesons.
The dashed line is for $\ep$ = 0.05
and the continuous line is for $\ep$=0.1.

\section{Comments and Comparisons}
We feel it is worthwhile to compare our results with those in a concurrent paper 
\cite{sd:meson}
that appeared as we were completing the manuscript for this paper.
There are a few noteworthy technical differences between the two.
Firstly, the treatment
of the longitudinal dynamics are different (DLCQ in Ref. \cite{sd:meson} while we
employed continuous basis functions). Secondly, the
pure glue string tension are different ($\sim \frac{2}{3}\, fm$ vs. $\sim 0.5\, fm$) 
Otherwise, both this work and Ref. \cite{sd:meson} were based on the
formalism outlined in Ref. \cite{mb:hala}.

We used continuous basis functions because there are some unresolved
issues regarding the convergence of DLCQ calculations in terms of the
DLCQ resolution \cite{me}. However, because the numerical approximations 
in both works are very crude, these issues are not yet very important and
therefore the main difference in this context is more in calculational 
details.

Because both works used the femtoworm approximation, the rms radius of
the $\pi$ is rather low compared to the experimental value. The lattice spacing, 
which defines the maximal $q\bar{q}$-separation,
is slightly larger in Ref. \cite{sd:meson}. Consequently, the resulting 
rms radius in Ref. \cite{sd:meson}, though still much smaller than the experimental result,
is slightly larger than ours.

Here we should also point out that the algebraic expression used in
Ref. \cite{sd:meson} for calculating $f_\pi$ is incorrect. 
With our normalization,
the result for $f_\pi$ in Ref. \cite{sd:meson} increases by a factor of
three \cite{private}
and becomes much more similar to ours.

Since both works determine the parameters in the effective theory by imposing
Lorentz symmetry as a nonperturbative renormalization condition, we feel that it is
important to have a handle on the violation of Lorentz invariance (or lack thereof) 
within the femtoworm approximation. That this violation is inevitable was first pointed 
out in 
Ref. \cite{taitung} (in which some results from this work were already presented)
and has been further
reiterated in section III of this paper. Since it is impossible to achieve both Lorentz 
invariance and the correct $\pi-\rho$ splitting simultaneously, one, therefore,
needs to have a quantitative idea of the degree to which Lorentz invariance
can be achieved while simlutaneously yielding the correct $\pi-\rho$ splitting.
Although Ref. \cite{sd:meson} provides several $\chi^2$ plots, 
it is not quantitatively clear from these plots how large the violations
of Lorentz invariance are for the $\pi-\rho$ dispersion relations at the
minimum of $\chi^2$. It is therefore not possible for us to compare the two works with
resepect to the degree to which Lorentz symmetry is violated.
Here we should also point out an interesting connection between
this work and Ref. \cite{sd:meson} on the one hand and Ref. \cite{mb:hala}
on the other: although the spirit of Ref. \cite{mb:hala} was more
phenomenological\footnote{The quark mass in 
the three particle Fock component in Ref. \cite{mb:hala}
was simply taken to be a constituent mass rather than being fitted to
optimize Lorentz symmetry and the ratio of the flip to nonflip coupling
was simply kept fixed in the fit of mass spectra.}, the final results
(parton distribution functions, distribution amplitudes)
obtained in Ref. \cite{mb:hala} were very similar
to both Ref. \cite{sd:meson} and this work.

Another difference between our results and those in Ref. \cite{sd:meson}
is that our $\pi$ distribution amplitude is closer to the 
`asymptotic' shape than the one in Ref. \cite{sd:meson} --- a behavior which is
also reflected in the parton distribution function. 
However, this difference is not really significant and it may just be
due to the differences in the handling of the longitudinal dynamics which
may favor slightly different sets of parameters.

\section{Summary and Conclusions}
We presented results from a study of light mesons in QCD formulated on a
transverse lattice using light-front quantization. In transverse lattice
QCD the gauge field in the (continuous) longitudinal directions is
described by a non-compact vector potential, and in the (discrete)
transverse directions by a non-compact link field.
Our approach was based on the color-dielectric formulation \cite{brett},
where one assumes that the fields have been ``smeared'' over a finite area
and thus are no longer subject to a strict $SU_C(N)$ constraint.

The smeared degrees of freedom are subject to effective interactions
which reflect the dynamics of the underlying microscopic degrees of
freedom. Although this effective interaction is not known {\it a priori},
it is highly constrained because it must still be gauge invariant
and it should still give rise to Lorentz invariant observables.
In this work we investigated transverse lattice QCD with fermions.
The starting point was an ansatz for the effective interactions
which contained interaction terms up to fourth order in the fields which
are still invariant under those symmetries not broken by the transverse
lattice.
The goal of this work was to determine the coefficients in this ansatz
by demanding enhanced Lorentz symmetry from physical observables.
As input parameters we used the physical masses of the $\pi$ and the $\rho$
as well as the string tension. What we found is that the splitting between
the $\pi$ and $\rho$ mesons strongly constrains the quark helicity flip
interaction term in the LF-Hamiltonian. Meson propagation in the $\perp$
direction involves helicity flip and nonflip hopping terms for the quarks.
Once the nonflip hopping term was constrained by fitting the observed
$\pi-\rho$  splitting, the only remaining parameter on which the physical
lattice spacing (as determined from the meson dispersion relation) depends
very strongly is the helicity nonflip term. Even though we scanned the
parameter
space thoroughly, we were only able to find matching values for
$a_\perp$ for the $\pi$ and $\rho$ mesons within about a 10\% - 20\% error.
Similar Lorentz invariance violations were found in the splitting between
$\rho$ mesons of helicity zero and one. We believe that both these
errors are due to the truncation of the Fock space
(which may contain an infinite number of particles)
as well as the ansatz for the effective interaction (which may contain an
infinite number of terms). In fact, given these rather crude approximations,
it is actually remarkable that the violations of Lorentz invariance in the
light meson sector are only about 10-20\% .
The model seems very promising though it displays some basic limitations as
far as restoration
of full Lorentz invariance is concerned. This problem was first identified
in one of our
earlier papers \cite{sudip} but subsequent new understanding \cite{us:meson}
about the couplings helped
isolate the problem and reduce it significantly for the low-lying mesons.
For the higher
mesons, the femtoworm approximation is not expected to work unless more link
field degrees
of freedom are introduced, a task that is possible but computationally
non-trivial.

While we were completing this work, similar studies of mesons on the
transverse lattice appeared  \cite{sd:hd,sd:meson}, which were based on DLCQ rather
than basis function techniques for handling the longitudinal
dynamics. However, there is another important difference between our work and
Ref. \cite{sd:hd}: we took the numerical value for the longitudinal
gauge coupling $G$ that was obtained in pure glue studies on the 
$\perp$ lattice \cite{brett}, while Ref.  \cite{sd:hd} allowed $G$ to 
`float' in addition
to the other parameters in order to restore Lorentz symmetry. As we 
explained in Section III, increasing $G$, as compared to other dimensionful
parameters, tends to decrease violations of Lorentz symmetry in the
$\pi/\rho$ dispersion relations. Therefore, the results presented in
Ref. \cite{sd:hd} where extrapolated from calculations performed with values of 
$G$ that were much larger than ours
(and larger than $G$ used in the pure glue calculation in Ref. \cite{brett}).
This difference in {\it procedure} explains to some extent the difference
in Results between this work and Refs. \cite{sd:hd}: most importantly, if
one increases $G$ in physical units, while keeping the
physical pion mass fixed, tends to decrease all other dimensionful parameters
\footnote{One can understand this fact by comparing with the simpler but
similar 1+1 dimensional 't Hooft model}.
The pion distribution amplitudes presented 
in Ref. \cite{sd:hd}, where $G$ was allowed to float completely freely, although
slightly `double humped', is actually numerically consistent with an almost
completely flat distribution since it was obtained by extrapolation
in the DLCQ parameter and there is some uncertainty in this procedure, and
this is consistent with the above observation regarding $G$. 
In Ref. \cite{sd:meson}, $G$ was no longer allowed to float and the results for 
the distribution amplitude are much more similar to ours.

\noindent {\bf Acknowledgments}
M.B. would like to thank S. Dalley for
many interesting and clarifying discussions about the $\perp$ lattice.
This work was supported by a grant from DOE (FG03-95ER40965) and through
Jefferson Lab by contract DE-AC05-84ER40150 under which the Southeastern
Universities Research Association (SURA) operates the Thomas Jefferson
National Accelerator Facility.

\appendix
\section{Overlap Integrals}
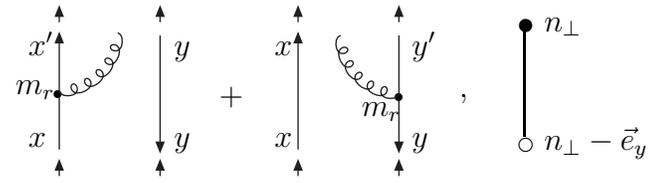
\begin{figure}
\begin{picture}(0,110)(280,10)

\put(392,46){$x$}
\put(447,46){$y$}
\LongArrow(402,35)(402,40)
\LongArrow(440,35)(440,40)

\LongArrow(402,45)(402,88)
\LongArrow(440,88)(440,45)

\LongArrow(402,93)(402,98)
\LongArrow(440,93)(440,98)
\put(392,81){$x'$}
\put(447,81){$y$}

\put(387,66){$m_r$}
\put(403,66){\circle*{3}}
\GlueArc(402,89)(22,-90,0){2}{5}

\put(464,62){+}

\put(485,46){$x$}
\put(537,46){$y$}
\LongArrow(492,35)(492,40)
\LongArrow(530,35)(530,40)

\LongArrow(492,45)(492,88)
\LongArrow(530,88)(530,45)

\LongArrow(492,93)(492,98)
\LongArrow(530,93)(530,98)
\put(485,81){$x$}
\put(537,81){$y'$}

\put(518,58){$m_r$}
\put(532,65){\circle*{3}}
\GlueArc(530,88)(22,180,270){2}{5}

\put(555,65){,}

\put(580,47){\circle{5}}
\put(579.5,50){\line(0,1){42}}
\put(580,92){\circle*{5}}
\put(587,45){$n_{\perp}-\vec{e}_y$}
\put(587,90){$n_{\perp}$}

\end{picture}
\caption{Spin non-flip hopping.}
\label{fig:non-flip}
\end{figure}

To understand the overlap integrals, let us look at two specific elements of the table (see
appendix B), namely, the first two elements of the first row. For the first element, clearly
there is no spin flip and the hopping is along the $y$-direction. There are two possibilities
through which this transition can take place. Pictorially, they are as shown in 
Fig. \ref{fig:non-flip}.

Notice that both diagrams yield the same 3-sector lattice configuration which is shown 
to the right. These two diagrams correspond to the terms containing 
$b^\d_\uparrow(\vec{n}_\perp,p_f^+)b_\uparrow(\vec{n}_\perp-\vec{e}_y,p_i^+)$ and 
$d^\d_\uparrow(\vec{n}_\perp,p_f^+)d_\uparrow(\vec{n}_\perp+\vec{e}_y,p_i^+)$ in the 
the spin non-flip interaction Hamiltonian. When these two terms are sandwiched between
$\mid qq\rangle$ and $\mid qqg\rangle$ from eqns. 2.2 and 2.3, one obtains the first matrix element 
of the table. In particular, $F^r_q$ and $F^r_{\bar{q}}$ are obtained when the integrals over
the quark and the antiquark momenta are carried out in the above matrix element as follows:

\bea
F_q^r = \int dx \int dy \psi^*(x,y) &&\left(\frac{m_r}{x}
+ \frac{m_r}{1-y}\right) \nonumber\\
&&\frac{1}{\sqrt{1-x-y}}\psi(1-y)
\eea
\bea
F_{\bar{q}}^r = \int dx \int dy \psi^*(x,y)&& \left(\frac{m_r}{y}
+ \frac{m_r}{1-x}\right) \nonumber\\
&&\frac{1}{\sqrt{1-x-y}}\psi(x)
\eea

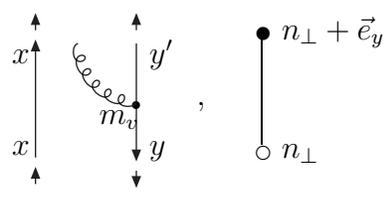
\begin{figure}
\begin{picture}(0,110)(340,10)

\put(485,46){$x$}
\put(537,46){$y$}
\LongArrow(492,35)(492,40)
\LongArrow(530,40)(530,35)

\LongArrow(492,45)(492,88)
\LongArrow(530,88)(530,45)

\LongArrow(492,93)(492,98)
\LongArrow(530,93)(530,98)
\put(485,81){$x$}
\put(537,81){$y'$}

\put(518,58){$m_v$}
\put(532,65){\circle*{3}}
\GlueArc(530,88)(22,180,270){2}{5}

\put(555,65){,}

\put(580,47){\circle{5}}
\put(579.5,50){\line(0,1){42}}
\put(580,92){\circle*{5}}
\put(587,45){$n_{\perp}$}
\put(587,90){$n_{\perp}+\vec{e}_y$}

\end{picture}
\caption{Spin flip hopping.}
\label{fig:flip}
\end{figure}

Similarly, let us
look at the second element of the first row of the table. Clearly, this is a spin
flip interaction and the hopping takes place in the $y$-direction. Pictorially, this is
shown in Fig. \ref{fig:flip}.

This diagram corresponds to the term in $\delta H^{flip}_y$ containing 
$d^\d_\uparrow(\vec{n}_\perp,p_f^+)d_\downarrow(\vec{n}_\perp+\vec{e}_y,p_i^+)$. 
The equivalent 3-sector lattice configuration is once again shown to the right.
The corresponding
matrix element is therefore obtained by sandwiching this term of the Hamiltonian between 
$\mid qq\rangle$ and $\mid qqg\rangle$ and in particular $f_{\bar{q}}$ is obtained when the integrals
over the quark and the antiquark momenta are carried out in the above matrix element as follows:

\bea
f_q = \int dx \int dy \psi^*(x,y) &&\left(\frac{m_v}{x}
- \frac{m_v}{1-y}\right) \nonumber\\
&&\frac{1}{\sqrt{1-x-y}}\psi(1-y)
\eea
\bea
f_{\bar{q}} = \int dx \int dy \psi^*(x,y)&& \left(\frac{m_v}{y}
- \frac{m_v}{1-x}\right) \nonumber\\
&&\frac{1}{\sqrt{1-x-y}}\psi(x).
\eea

%\clearpage
\section{Transition Matrix}

\begin{eqnarray*}
\begin{array}{|c||c|c|c|c|} \hline
\mbox{states}
& \mid\psi_2,\uparrow\uparrow\rangle
& \mid\psi_2,\uparrow\downarrow\rangle
& \mid\psi_2,\downarrow\uparrow\rangle
& \mid\psi_2,\downarrow\downarrow\rangle \\ \hline\hline

&&&& \\
\mid\psi_3,\uparrow\uparrow,+\vec{e}_y\rangle
& F^r_qe^{-\frac{ik_y}{2}}+F^r_{\bar{q}}e^{\frac{ik_y}{2}}
& -f_{\bar{q}}e^{\frac{ik_y}{2}}
& -f_{q}e^{-\frac{ik_y}{2}}
& 0 \\
&&&& \\ \hline

&&&& \\
\mid\psi_3,\uparrow\downarrow,+\vec{e}_y\rangle
& -f_{\bar{q}}e^{\frac{ik_y}{2}}
& F^r_qe^{-\frac{ik_y}{2}}+F^r_{\bar{q}}e^{\frac{ik_y}{2}}
& 0
&  -f_{q}e^{-\frac{ik_y}{2}} \\
&&&& \\ \hline

&&&& \\
\mid\psi_3,\downarrow\uparrow,+\vec{e}_y\rangle
& -f_qe^{-\frac{ik_y}{2}}
& 0
& F^r_qe^{-\frac{ik_y}{2}}+F^r_{\bar{q}}e^{\frac{ik_y}{2}}
&  -f_{\bar{q}}e^{\frac{ik_y}{2}} \\
&&&& \\ \hline

&&&& \\
\mid\psi_3,\downarrow\downarrow,+\vec{e}_y\rangle
& 0
&  -f_qe^{-\frac{ik_y}{2}}
& -f_{\bar{q}}e^{\frac{ik_y}{2}}
& F^r_qe^{-\frac{ik_y}{2}}+F^r_{\bar{q}}e^{\frac{ik_y}{2}}\\
&&&& \\ \hline

&&&& \\
\mid\psi_3,\uparrow\uparrow,+\vec{e}_x\rangle
& F^r_qe^{-\frac{ik_x}{2}}+F^r_{\bar{q}}e^{\frac{ik_x}{2}}
&  -if_{\bar{q}}e^{\frac{ik_x}{2}}
& -if_qe^{-\frac{ik_x}{2}}
& 0 \\
&&&& \\ \hline

&&&& \\
\mid\psi_3,\uparrow\downarrow,+\vec{e}_x\rangle
& if_{\bar{q}}e^{\frac{ik_x}{2}}
& F^r_qe^{-\frac{ik_x}{2}}+F^r_{\bar{q}}e^{\frac{ik_x}{2}}
& 0
& -if_qe^{-\frac{ik_x}{2}} \\
&&&& \\ \hline

&&&& \\
\mid\psi_3,\downarrow\uparrow,+\vec{e}_x\rangle
& if_qe^{-\frac{ik_x}{2}}
& 0
& F^r_qe^{-\frac{ik_x}{2}}+F^r_{\bar{q}}e^{\frac{ik_x}{2}}
& -if_{\bar{q}}e^{\frac{ik_x}{2}} \\
&&&& \\ \hline

&&&& \\
\mid\psi_3,\downarrow\downarrow,+\vec{e}_x\rangle
& 0
& if_qe^{-\frac{ik_x}{2}}
& if_{\bar{q}}e^{\frac{ik_x}{2}}
& F^r_qe^{-\frac{ik_x}{2}}+F^r_{\bar{q}}e^{\frac{ik_x}{2}}  \\
&&&& \\ \hline
\end{array}
\end{eqnarray*}

\newpage

\begin{eqnarray*}
\begin{array}{|c||c|c|c|c|} \hline
\mbox{states}
& \mid\psi_2,\uparrow\uparrow\rangle
& \mid\psi_2,\uparrow\downarrow\rangle
& \mid\psi_2,\downarrow\uparrow\rangle
& \mid\psi_2,\downarrow\downarrow\rangle \\ \hline\hline

&&&& \\
\mid\psi_3,\uparrow\uparrow,-\vec{e}_y\rangle
& F^r_qe^{\frac{ik_y}{2}}+F^r_{\bar{q}}e^{-\frac{ik_y}{2}}
& f_{\bar{q}}e^{-\frac{ik_y}{2}}
& f_qe^{\frac{ik_y}{2}}
& 0 \\
&&&& \\ \hline

&&&& \\
\mid\psi_3,\uparrow\downarrow,-\vec{e}_y\rangle
& f_{\bar{q}}e^{-\frac{ik_y}{2}}
& F^r_qe^{\frac{ik_y}{2}}+F^r_{\bar{q}}e^{-\frac{ik_y}{2}}
& 0
& f_qe^{\frac{ik_y}{2}}  \\
&&&& \\ \hline

&&&& \\
\mid\psi_3,\downarrow\uparrow,-\vec{e}_y\rangle
& f_qe^{\frac{ik_y}{2}}
& 0
& F^r_qe^{\frac{ik_y}{2}}+F^r_{\bar{q}}e^{-\frac{ik_y}{2}}
& f_{\bar{q}}e^{-\frac{ik_y}{2}} \\
&&&& \\ \hline

&&&& \\
\mid\psi_3,\downarrow\downarrow,-\vec{e}_y\rangle
& 0
& f_qe^{\frac{ik_y}{2}}
& f_{\bar{q}}e^{-\frac{ik_y}{2}}
& F^r_qe^{\frac{ik_y}{2}}+F^r_{\bar{q}}e^{-\frac{ik_y}{2}}  \\
&&&& \\ \hline

&&&& \\
\mid\psi_3,\uparrow\uparrow,-\vec{e}_x\rangle
& F^r_qe^{\frac{ik_x}{2}}+F^r_{\bar{q}}e^{-\frac{ik_x}{2}}
& if_{\bar{q}}e^{-\frac{ik_x}{2}}
& if_{q}e^{\frac{ik_x}{2}}
& 0 \\
&&&& \\ \hline

&&&& \\
\mid\psi_3,\uparrow\downarrow,-\vec{e}_x\rangle
& -if_{\bar{q}}e^{-\frac{ik_x}{2}}
& F^r_qe^{\frac{ik_x}{2}}+F^r_{\bar{q}}e^{-\frac{ik_x}{2}}
& 0
& if_{q}e^{\frac{ik_x}{2}} \\
&&&& \\ \hline

&&&& \\
\mid\psi_3,\downarrow\uparrow,-\vec{e}_x\rangle
& -if_qe^{\frac{ik_x}{2}}
& 0
& F^r_qe^{\frac{ik_x}{2}}+F^r_{\bar{q}}e^{-\frac{ik_x}{2}}
& if_{\bar{q}}e^{-\frac{ik_x}{2}} \\
&&&& \\ \hline

&&&& \\
\mid\psi_3,\downarrow\downarrow,-\vec{e}_x\rangle
& 0
& -if_qe^{\frac{ik_x}{2}}
& -if_{\bar{q}}e^{-\frac{ik_x}{2}}
& F^r_qe^{\frac{ik_x}{2}}+F^r_{\bar{q}}e^{-\frac{ik_x}{2}} \\
&&&& \\ \hline
\end{array}
\end{eqnarray*}

\newpage

\newpage

\newpage

\newpage

\newpage

\newpage

\newpage

\begin{figure}
\unitlength1.cm
\begin{picture}(15,6.5)(-11.5,0.5)
\includegraphics{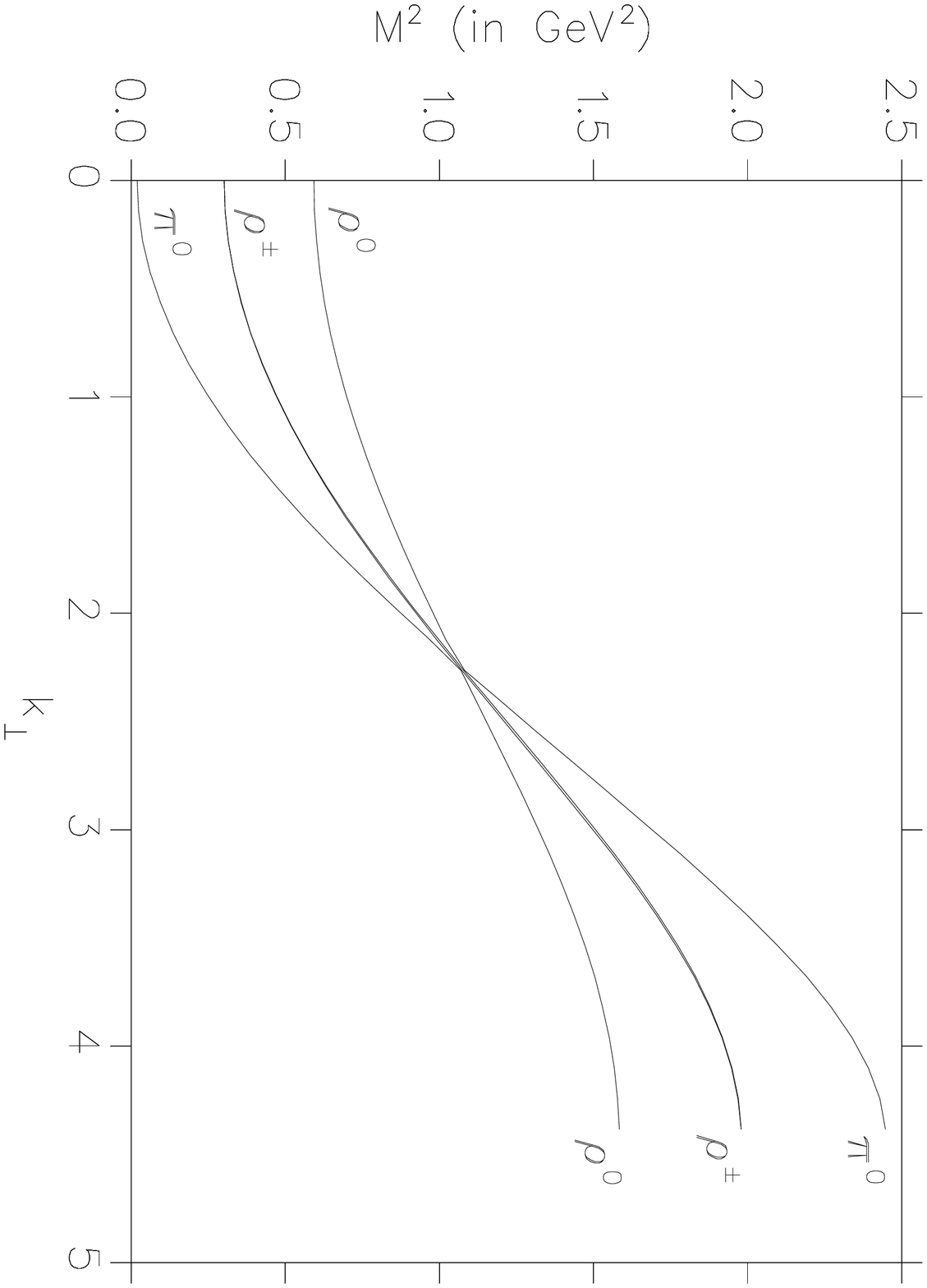}
\end{picture}
%\caption{The diagonal dispersion relations for the $\pi$ and $\rho$ mesons.}
%\label{fig:bzoned}
\end{figure}

\newpage

\newpage

\newpage

\newpage

\end{document}